\documentclass[twocolumn]{webofc}

\usepackage[varg]{txfonts}   
\usepackage{hyperref}
\usepackage{url}

\hypersetup{colorlinks=true,citecolor=blue,urlcolor=blue,linkcolor=blue}

\usepackage{scalerel}
\newlength\curht

\def\defaultstartht{14pt}
\newcommand\diminish[2][\defaultdimfrac]{%
  \curht=\defaultstartht\relax
  \def\dimfrac{#1}%
  \diminishhelpA{#2}%
}
\newcommand\diminishhelpA[1]{%
  \expandafter\diminishhelpB#1\relax%
}
\def\diminishhelpB#1#2\relax{%
  \scaleto{\strut#1}{\curht}%
  \curht=\dimfrac\curht\relax%
  \ifx\relax#2\relax\else\diminishhelpA{#2}\fi%
}
\def\well{1.0594630943592953098431053149397484958171844482421875}

\usepackage{ifthen}
\newcommand{\linkAudio}[2][]{%
\ifthenelse{\equal{#1}{}}{\href{https://gitup.uni-potsdam.de/henkel/sonification_cd25/-/blob/main/Public_mp3s/#2}{\texttt{\small #2}}}%
{\href{https://gitup.uni-potsdam.de/henkel/sonification_cd25/-/blob/main/Public_mp3s/#1}{\texttt{\small #2}}}}

\renewcommand{\doiwoc}[1]{%
{\small\texttt{\href{https://doi.org/#1}{DOI:#1}}}%
}

\newcommand{\ch}[1]{{#1}}

\usepackage[normalem]{ulem}
\renewcommand{\sout}[1]{}

\usepackage{fancyhdr}
\begin{document}


\title{Quantum Listenings}
%
%
\subtitle{\protect\sout{Amateur}Sonification of Vacuum and other Noises}

\author{\firstname{Carsten} \lastname{Henkel}\inst{1}
\fnsep\thanks{\email{carsten.henkel@uni-potsdam.de}}}

\institute{Universität Potsdam, Institut für Physik und Astronomie, Germany}

\abstract{%
The sensory perceptions of vision and sound may be considered 
as complementary doorways towards interpreting and understanding
physical phenomena. 
We provide a few selected samples where scientific data of systems
usually not directly accessible to humans may be listened to. The
examples are chosen close to the regime where quantum mechanics is applicable. 
Visual and auditory renderings are compared with some connections to
music, illustrating in particular a kind of fractal complexity along 
the time axis.
}
%
%
\maketitle

\section{Introduction}
\label{intro}

``Do you see this point? -- No, I do not understand.'' 
This fictitious sample of a dispute points to 
the conceptual links between sensory perception (seeing, hearing) 
and logical or physical explanations which may be the key achievements 
of the natural sciences.
In this paper, I would like to discuss the complementarity 
in illustrating physical phenomena with two-dimensional graphics
and acoustic signals, with the aim of comparing 
ways of understanding.
A picture may be captured at a glance, while listening to a sentence
develops in time. In particular perceiving and memorizing sound seems
to proceed in a fractal way as time unfolds. 
This is in part due to the physical processes
of sound generation like the many attributes of a tone in musical 
acoustics (spectral color, pitch, attack, fading \ldots).
On the other hand, the brain seems to function by compressing acoustic 
information in a formidable way, building on recognizable ``molecular 
unit'' patterns and then organising them in a hierarchical way from
phrases to epochs in the history of music.

In this paper, we review a few basic properties of perceiving and
encoding sound (Sec.\:\ref{s:sound-perception}). We then give examples
for mapping observables from the atomic or molecular scale to sound
(Secs.\:\ref{s:quantum-chord}, \ref{s:listen-to-AFM}). 
In Secs.\:\ref{s:noise-examples}, \ref{s:Bose-gas-spectrogram},
spectra of quantum noise are discussed and converted into the audio range.
Next to the examples, \sout{we}\ch{a} few exercises are suggested for further 
applications.

This material is provided by an amateur of sonification and is by no
means meant to be comprehensive.
\ch{Useful resources can be found via the conference series on
auditory display and its web site \cite{ICAD}.}
The chosen examples hopefully help to appreciate the many facets of complexity
and disorder in the physical world.
Audio files are available from the repository
\href{https://gitup.uni-potsdam.de/henkel/sonification\_cd25}{\texttt{\small gitup.uni-potsdam.de/henkel/sonification\_cd25}}
\cite{gitup_repo}.

\section{Basics of Sonification}
\label{s:basics}


\subsection{Human sound perception}
\label{s:sound-perception}

The sensorial apparatus for auditory perception of a young person spans
the frequency range of 20\,Hz--20\,kHz. 
This is about $2^{10}$, i.e., ten octaves, much wider than the single octave
of the visible spectrum (between $c/780\,{\rm nm}$ and $c/380\,{\rm nm}$).
Frequencies below 10\,Hz are perceived in the time domain via the onset
and fading of a sound. This illustrates an interesting ``separation of
time scales'' that provides the basis for mapping one-dimensional sound
data into a two-dimensional time-frequency representation 
(see Sec.\:\ref{s:noise-basics}). The inner ear is performing a spectral
analysis over the mentioned frequency range on times scales faster than
100\,msec, as listeners trained to musical samples may testify.
%

\begin{figure}[htbp]
   \centering
   \includegraphics[width=0.9\columnwidth]{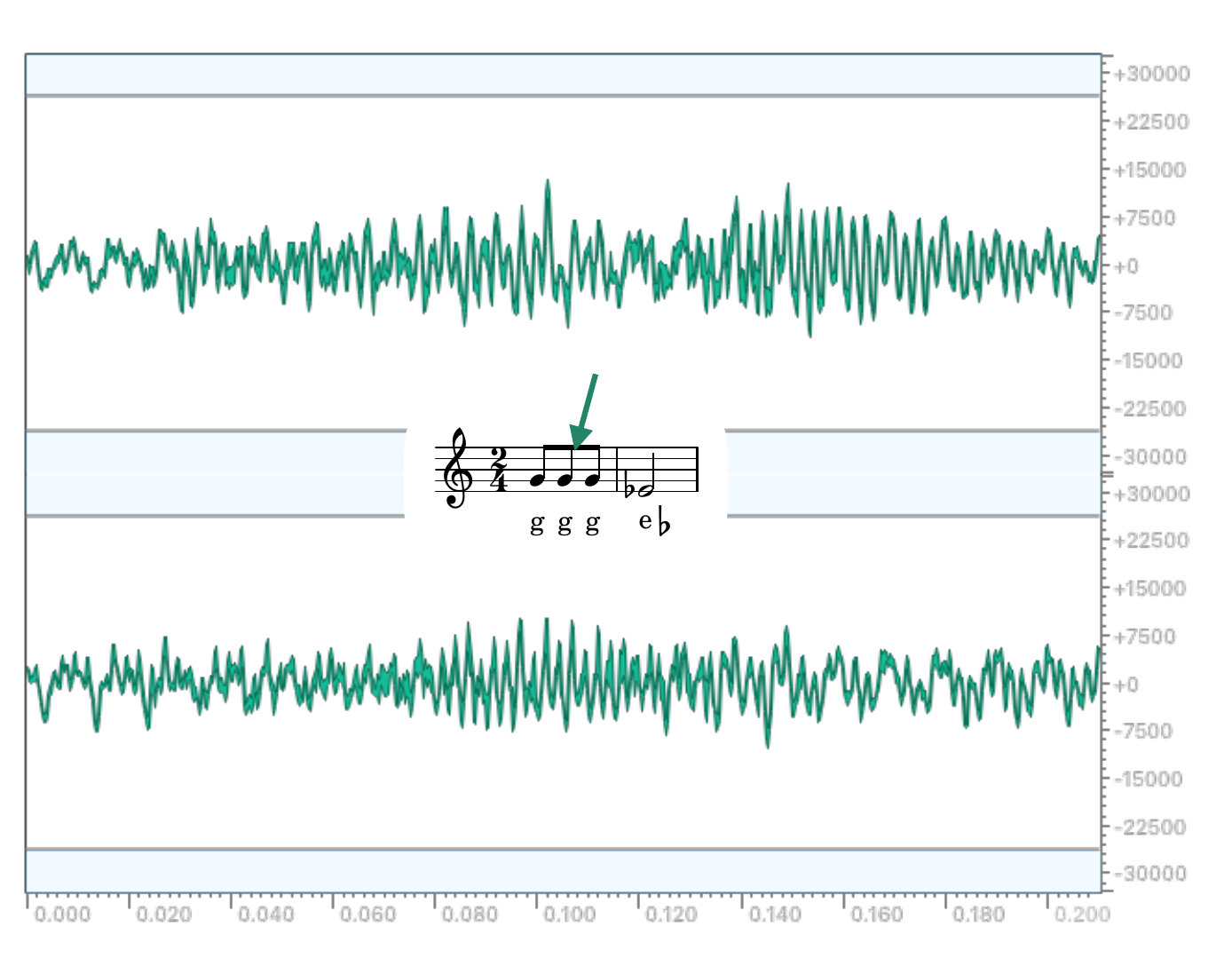}
   \caption[]{Example of an acoustic waveform (two stereo channels are shown). 
   These 210\,ms represent the 2nd eighth note (see insert,
   g$^1 \approx 392\,{\rm Hz}$, i.e.~period 2.55\,ms) 
   of Beethoven's 5th Symphony,
   1st Movement (c minor, op.~67)
   \cite{live-in-Ramallah}.
   Audio file: \linkAudio{Beethoven5\_2nd-8th.mp3}.}
   \label{fig:Beethoven5-2nd-8th}
\end{figure}

Modern sound treatment is done in a discrete (digital) representation
where an acoustic signal is a sequence $u(t_i)$ of integers in the
range 
$-2^{15} \le u < 2^{15} = 32\,768$ (16 bit depth amplitude)
and the elementary time step is ${\rm d}t = 1/44.1\,{\rm kHz} = 22.7\,\text{µs}$.
%
The physical properties of the signal are often quantified using 
logarithmic scales,
for example the familiar decibel (dB) unit for loudness. 
Since you may easily turn down the volume, we focus here on typical scales 
in the time and frequency domains.
%
An example is provided in Fig.\:\ref{fig:Beethoven5-2nd-8th} where the
stereo channels over a duration of 200\,msec for a single note are shown. 
The eye barely makes sense of these erratic signals, 
but try to listen to this sample (see caption). 
Although it covers less than 100 periods
of the fundamental vibration (and less than $10^4$ data points per channel), 
we seem to recognise that this was played by a 
string orchestra (!).
%

\subsection{Mapping to sound}
\label{s:signals-to-sound}

\begin{figure}[tbph]
\centering
\includegraphics[width=0.27\textwidth]{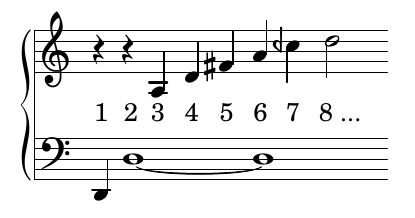}
\caption[]{Harmonics of D ($\approx 73\,{\rm Hz}$) up to the eighth (three octaves). 
In the usual tuning, the 6th harmonic appears at a$^1 = 440\,{\rm Hz}$. 
The 7th harmonic
appears approximately a quarter tone below c$^2 = 523\,{\rm Hz}$, 
i.e., at $\approx 508\,{\rm Hz}$ (accidental \reflectbox{$\flat$}),
\emph{cf.}\ audio file
\linkAudio{obertonreihe.mp3}.
(In diatonic tuning, D would appear at $440/6 = 73.333\,{\rm Hz}$,
in well-tempered tuning, at $2^{-7/12} \, 440/4 = 73.416\,{\rm Hz}$.)
}
\label{fig:obertoene}
\end{figure}

To illustrate frequencies,
the piano keyboard or musical notation in general is 
a perhaps familiar example. 
Tones in music are organised in a logarithmic scale, 
each factor of two being represented by a fixed distance (one octave).
A linear progression of frequencies, naturally provided by the
harmonics of a periodic signal, thus clusters into narrower intervals
in the musical sense, as the number of harmonics grows beyond eight
(Fig.\:\ref{fig:obertoene}). 
The seventh harmonic already falls between the discrete notes used in
western music, and a quarter tone only provides an approximation.
As musical notes increase in pitch, their absolute frequency spacing 
hence increases,
which is not much different from the distribution of prime numbers
(first studied by Gauß and Legendre). 
%

The fundamental musical identification of tones separated
by an octave introduces a periodicity along the (logarithmic) frequency axis.
The lettering of notes in music is a prime example. 
Variants of this ``coding'' can be used for ``mathematical compositions''
\ch{in the spirit of Iannis Xenakis \cite{Serra93}.}
One brief illustration is provided
in Fig.\:\ref{fig:pi-over-e-thumb} (see also 
Figs.\:11--13 in the Appendix). 
In hommage to J. S. Bach and A. Schönberg, 
the well-tempered chromatic scale with its twelve evenly spaced notes
(ratio $2^{1/12}$),
is used to represent the digits $0, 1 \ldots 9, 10=a, 11=b$ in base 12. 
Taking the digits $\pi$ as a never-ending resource, 
one gets the upper voice of the score in Fig.\:\ref{fig:pi-over-e-thumb}. 
The lower voice is computed from Euler's number $e$, 
but represented in base 5.
Its digits are converted into another musical set of notes, the
pentatonic scale.
We take the same subset as the incipit of M. Mussorgsky's
``Promenade'' from \emph{Pictures of an Exhibition}.
The audio file \linkAudio{pi\_over\_e\_classical.mp3} and other
variations can be downloaded from the repository 
mentioned in the
Introduction \cite{gitup_repo}.
In the ``$\pi/e$ walk'' (Fig.\:12 in the Appendix), 
the music is allowed to wander up and down beyond 
a single octave, similar to Brownian motion.

\begin{figure}[tbph]
\centering
\hspace*{-5mm}  
\includegraphics[width=1.1\columnwidth]{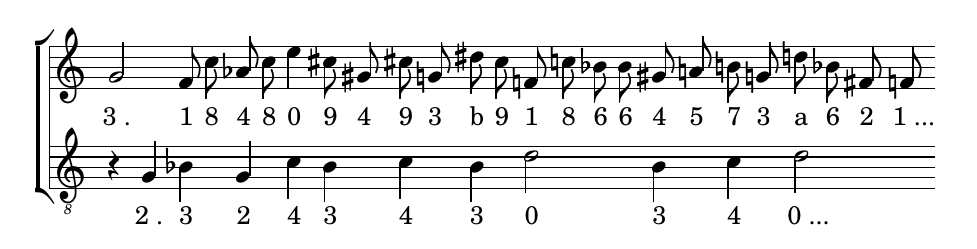}
\caption[]{Sonified duet of $\pi$ over ${\rm e}$. The digits 
of $\pi$ in base twelve are encoded into the well-tempered half-tones
from f$^1$ to e$^2$, 
while ${\rm e}$'s digits in base five are mapped to a pentatonic
scale. 
The only manual edits were applied to duration: 
the first note and the digit 0 getting longer durations.
For a longer version, see Fig.\:11 in the Appendix and audio 
file \linkAudio{pi\_over\_e\_classical.mp3}.}
\label{fig:pi-over-e-thumb}
\end{figure}

Such a piece may be perceived \emph{prima vista} as ``complex'',
and when listening, it sounds ``disordered'' or even ``pointless''.
Ear and brain, at least when building on classical western music education,
are missing ``structures''. These are expected to be fractal in time,
grouping tones into basic motifs, phrases, themes, and entire movements.
This melody, however, seems to crumble and decompose into its atoms. 
\ch{Each of the notes is highly conventional (at least in the violin/bassoon
chosen instrumentation), but taken as a sequence, they do not blend easily 
into larger musical structures.}
It does not come as a surprise that any ``interesting'' sequence of numbers
can be found in the digits 
of an irrational number like $\pi$ or e. 
And indeed, after a few listenings, also the ear starts to remember
(recognize) certain fragments, and be they the somehow artificially
longer notes of the 0 digit.
An example for an extended fragment that somehow ``makes (musical) sense''
is highlighted in Fig.\:11 of the Appendix (grey bars), corresponding
to the digits 1332 34411 23230 in the lower voice. This impression 
quite probably arises because the second group nearly repeats the first
one, bringing forward a connection between ``sense'' and ``lower
entropy''.
\ch{Indeed, both concepts fundamentally relate to the compression of data,
as illustrated by constructing entropy from Shannon's information theory 
concepts.}

\subsection{A simple quantum chord}
\label{s:quantum-chord}

More abstract examples of sound can be constructed 
based on the physics of vibrations like in electrodynamics 
and quantum mechanics. 
\ch{Nearly hundred years ago, when the first Raman spectra were 
recorded, Andrews mapped the lines found for substances like ethanol 
or gasoline into a set of music notes, as described in 
Ref.\:\cite{Watson_1931}.}
\sout{Indeed}Quoting one of the quantum founders,
A. Sommerfeld \cite{Sommerfeld_Atombau}:

``[Die Quantentheorie] ist das geheimnisvolle Or\-ganon,
auf dem die Natur die Spektralmusik spielt und nach dessen Rhythmus sie
den Bau der Atome und Kerne regelt.'' -- Quantum theory is the mysterious
organ which Nature is playing its spectral music on and whose rhythm
She uses to \ch{build and rule} atoms and nuclei. (Transl.~C.H.)

\begin{figure}[tbph]
   \centering
    \hspace*{-4mm}  
   \includegraphics[width=1.1\columnwidth, clip]{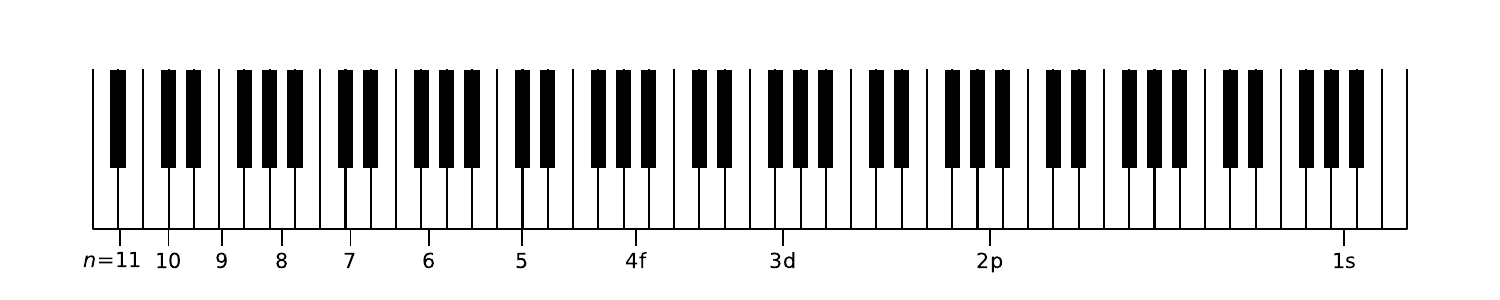}
   \caption[]{Frequencies corresponding to the binding energies of
   the Hydrogen atom mapped to a piano keyboard. Note the beautiful
   features of ``atomic harmony'' \cite{Sommerfeld_Atombau}:
	the levels $n = 2, 3$ are separated by two fifths
	(ratio $(2/3)^2$) and $3, 4$ by two fourths ($(3/4)^2$). 
	(The
	slight offset of the levels $n = 3, 6$ with respect
	to the keys is actually due to the well-tempered tuning assumed
	here where fifths and fourths are not ``pure''.)
	The audio file \linkAudio{hydrogen\_keys.mp3} renders the frequencies
	approximately on a well-tempered keyboard, including quarter tones.
   }
   \label{fig:hydrogen-piano}
\end{figure}

\ch{This idea has blossomed in recent years to produce the field 
of quantum sonification.
A few entry points into it can be found in Refs.\,\cite{Yamada23, Christie24, Blencowe25}.
If we turn to}
the simplest quantum system with a discrete spectrum, this is probably the
Hydrogen atom. Indeed, the energy levels of the harmonic oscillator 
have already
been illustrated by the harmonics of Fig.\:\ref{fig:obertoene}.
There are infinitely many bound levels in Hydrogen and other atoms due to 
the long-range Coulomb interaction. The usual representation is putting
the ground state at the bottom, less strongly bound levels appear above
it. 
In terms of frequencies, the reverse ordering may be more natural.
We thus take the binding energies $E_n = 1/2n^2\,{\rm a.u.}$ 
of the Balmer formula and translate them into a frequency in the 
acoustic range.
Restricting ourselves for simplicity to the 88 keys on the piano,
we are limited to a ratio from highest to lowest frequencies of
$2^{87/12} \approx 150$.
We map the $n = 1$ level (binding energy $13.6\,{\rm eV}$) 
to a very high key (a$^4 = 3520\,{\rm Hz}$)
and find that the lowest one that would still be playable on the piano 
is the Rydberg level $n = 11$ ($0.112\,{\rm eV}$),
see the key B$^\flat_2 \approx 29\,{\rm Hz}$ 
in Fig.\:\ref{fig:hydrogen-piano}.
An unusual aspect of this representation 
is the relatively wide spacing between large-$n$ levels, 
which arises, of course, from the logarithmic map 
to the musical keyboard.
--
As an intriguing \textbf{exercise}, look up a few low-lying levels
of the alkali atoms, say, and play the corresponding quantum (defect) 
chords.


How would the time scales due to the lifetimes of atomic
levels translate into this acoustic picture? 
We may represent a decaying state as a pure tone with a slowly
decreasing intensity.
After exciting the atom with a laser pulse to the Rydberg level 10p$_{3/2}$, say,
a deep vibration (the lowest C$\sharp_1$ key, $34.7\,{\rm Hz}$,
in Fig.\:\ref{fig:hydrogen-piano}) would signal the occupation of this state.
Its radiative decay is, however, very long compared
to the electronic oscillation period (binding frequency
$32.9\,{\rm THz}$ measured relative to the ionization potential)
\cite{NIST_database}.
The fastest decay channel is to the 1s state with an Einstein coefficient
$A = 4.21{\rm E}06/{\rm s}$,
and in the ``keyboard representation'', its actual decay time
would be more than two days (!).
Even the Lyman-$\alpha$ line (2p $\to$ 1s, 122\,nm) which has a much
shorter lifetime ($A = 6.265{\rm E}08/{\rm s}$) would have 2p ``vibrate'' 
for more than an hour.

An interesting question is whether the ear is able to 
perceive phase relations between harmonics. 
For quantum states, this is equivalent to telling apart a coherent
superposition from a statistical mixture. 
\ch{Conversely, stereo audio signals can be used
to bring coherences to sound, thanks to binaural hearing \cite{Christie24}.}
In any case, to provide an actual ensemble, one would have
to resort to ergodicity and perform a statistical analysis over some time
window. We come back to the corresponding autocorrelation functions in 
Sec.\:\ref{s:noise-basics}.

\subsection{Molecular vibrations}
\label{s:listen-to-AFM}

\begin{figure*}[htbp]
   \centering
   \begin{minipage}{0.4\textwidth}
   \mbox{(a) $d = 0.70\,{\rm nm}$}\\
   \includegraphics[width=06.5cm]{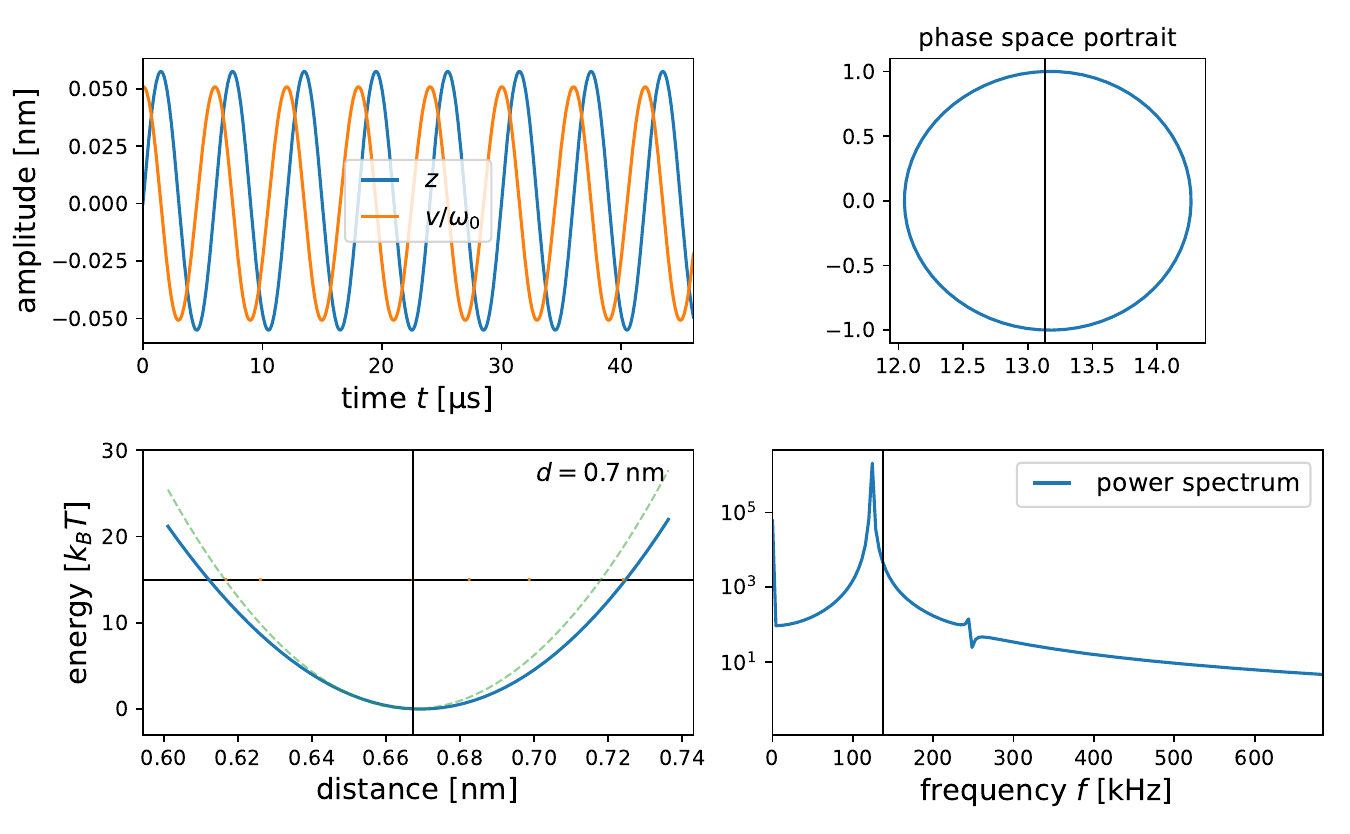}
   \end{minipage}
   \hspace*{5ex}
   \begin{minipage}{0.4\textwidth}
   \mbox{(b) $d = 0.61\,{\rm nm}$}\\
   \includegraphics[width=06.5cm]{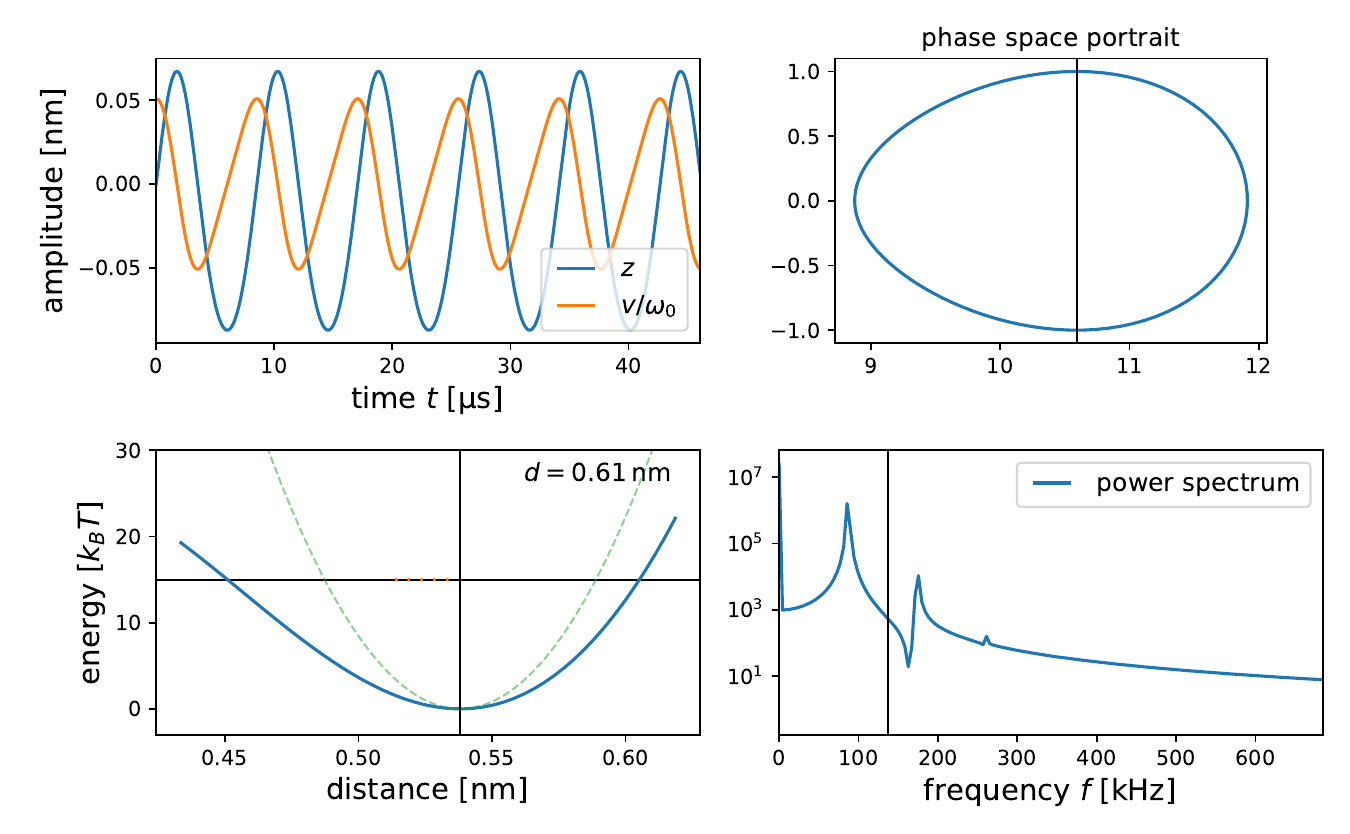}
   \end{minipage}
\\[2ex]
   \begin{minipage}{0.4\textwidth}
   \mbox{(c) $d = 0.60\,{\rm nm}$}\\
   \includegraphics[width=06.5cm]{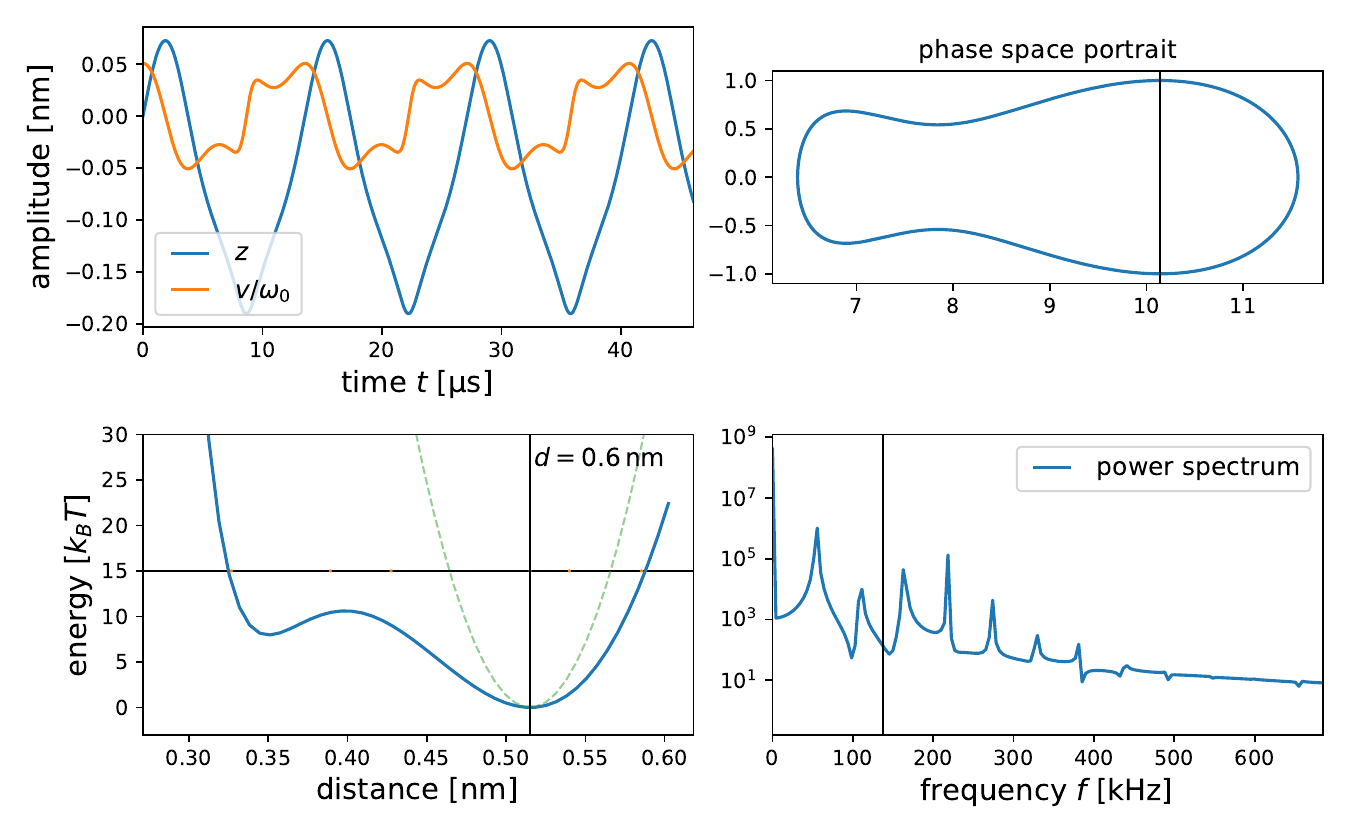}
   \end{minipage}
   \hspace*{5ex}
   \begin{minipage}{0.4\textwidth}
   \mbox{(d) $d = 0.57\,{\rm nm}$}\\
   \includegraphics[width=06.5cm]{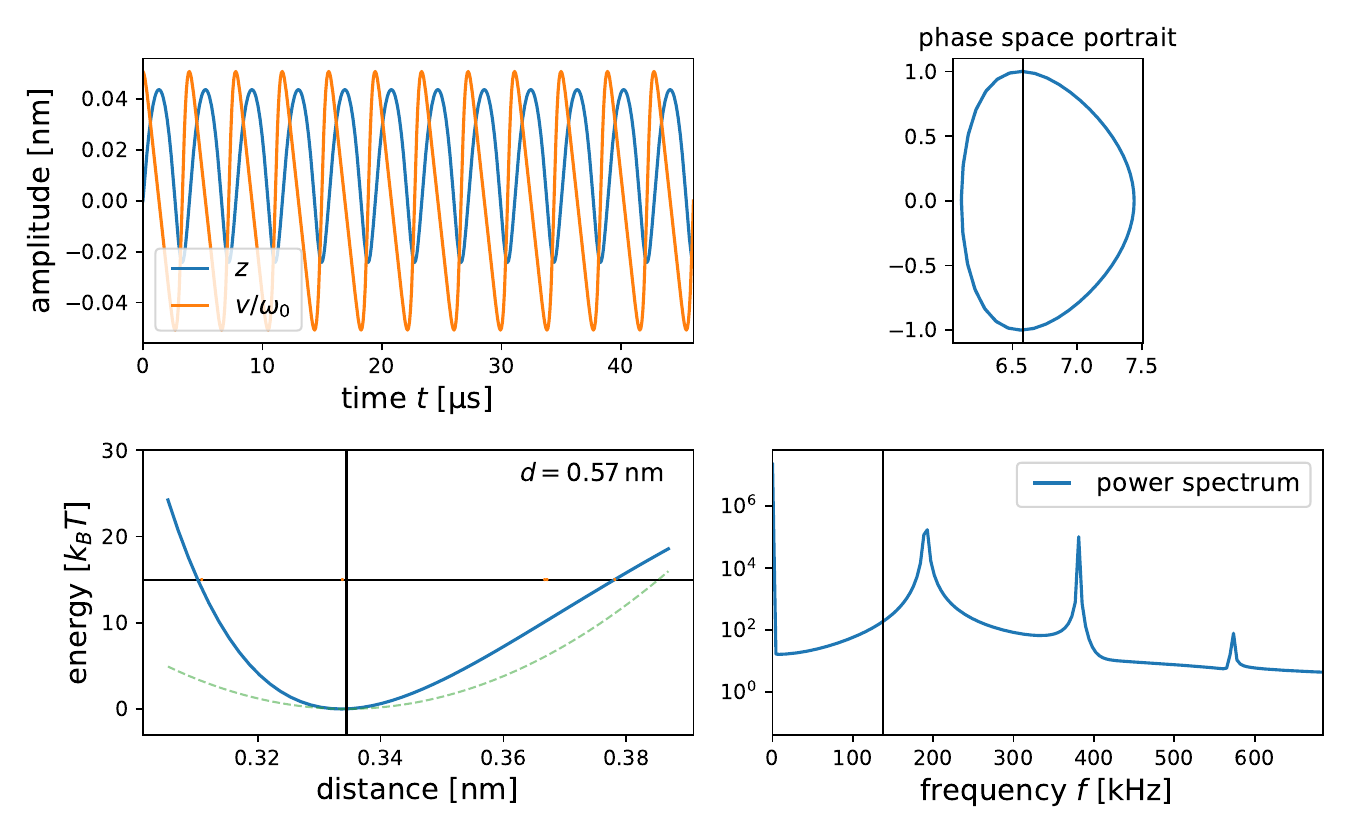}
   \end{minipage}
   \caption[]{Analysis of nonlinear oscillations upon
   approaching a cantilever to a solid surface. 
   In each of the four panels, one
   finds at top left: trajectory, top right:
   phase space protrait, bottom left: effective potential including
   a harmonic contribution from the piezo-control, bottom right:
   power spectrum (Fourier transform) of the vibration. 
   The oscillation occurs at a fixed energy about $15\,k_BT$
   above the potential minimum (horizontal line).
   The vertical
   line in the spectra 
   marks the eigenfrequency $\Omega/2\pi \approx 140\,{\rm kHz}$.
   In the \ch{acoustic version} (\linkAudio{listen\_approach\_curve.mp3}), 
   the time series has been re-scaled
   to that $\Omega/2\pi$ corresponds to roughly $400\,{\rm Hz}$.}
   \label{fig:listen-to-AFM}
\end{figure*}

Scanning microscopes are powerful instruments to image the world of atoms 
and molecules. In these devices, a sharp tip is mounted
on a flexible beam (cantilever) and is approached to a solid sample
using piezo-control stages. The cantilever can be modelled as
a weakly damped harmonic oscillator which is subject to a force
with a steep distance dependence:
\begin{equation}
m \left( \ddot{z} + \gamma \dot{z} + \Omega^2 (z - z_0) \right) =
- \frac{\partial V }{ \partial z} + f(t)
\,,
\label{eq:LJ-AFM-oscillator}
\end{equation}
where $m$ is an effective mass, 
$z_0$ the nominal equilibrium position,
and $f(t)$ an external drive. 
A typical interaction potential
is of the Lennard-Jones type
\begin{equation}
V(z) = - \frac{ c_3 }{ 3 z^3 } + \frac{ c_{9} }{ 9 z^{9} }
\,.
\label{eq:}
\end{equation}
At large distance, expanding the potential 
around the equilibrium position $z_0$, one gets
a complex dispersion relation
\begin{equation}
\omega^2 + {\rm i} \gamma \omega - \Omega^2 
+ \frac{ 4 c_3 }{ m z_0^5 } - \frac{ 10 c_{9} }{ m z_0^{11} } 
= 0
\,.
\label{eq:frequency-shift}
\end{equation}
Neglecting the last term, the long-range surface attraction
downshifts by its intrinsic curvature (second derivative) 
the cantilever vibration. 
This is a common method to measure surface forces
\cite{Giessibl97}, but requires some model assumptions to
reconstruct a full force-distance curve.
As the cantilever approaches the surface, the linearization
behind Eq.\,(\ref{eq:frequency-shift}) breaks down, and the
vibration becomes anharmonic. 
This is shown in Fig.\:\ref{fig:listen-to-AFM} where the numerical
solution to Eq.\,(\ref{eq:LJ-AFM-oscillator}) is plotted: the
upper left panels give the quadratures in the time domain,
the lower right ones the spectrum of the vibration.
Upon approaching a sample, eventually neither the natural spring constant
$k = m \Omega^2$ nor the piezo-control force are able to compensate
for the attractive forces: the cantilever then moves into a second
equilibrium position closer to the surface. Right at the transition,
the potential is bistable [around distance $0.60\,{\rm nm}$ 
in Fig.\:\ref{fig:listen-to-AFM}(c)]. Between this point and
$0.57\,{\rm nm}$, the down-shifted vibration disappears and
is replaced by a strongly coloured spectrum with many harmonics.

It is quite interesting to appreciate these features acoustically
from the provided \texttt{\small .mp3} file.
The sonification is generated from the numerical solution
with the \texttt{\small scipy.io.wavfile} package. 
The nominal frequency $\Omega/2\pi$ has been mapped to the
audio band. 
We have used audio software to remove phase jumps
that slipped into the simulated raw data and to patch datasets at
different distances into a single file.  
It is surprising that in
acoustic synthesis of this kind, the ear clearly perceives 
a single phase slip or discontinuity as a ``knack''. 
\ch{Actual experimental data also contain a noise floor 
due to amplifiers and the intrinsic thermal motion, as can
be heard in Ref.\:\cite{Blencowe25}.
A natural \textbf{exercise} to take this into account 
would be a stochastic model with an additional Langevin 
force on the rhs of Eq.\,(\ref{eq:LJ-AFM-oscillator}).
Non-additive noise is expected to emerge in the strongly
nonlinear regime close to contact and may lead to a qualitatively
different spectral density \cite{Biro_2005a}.}

This way of ``listening'' to an AFM may prove useful
for experimenters. The second and higher harmonics
in Fig.\:\ref{fig:listen-to-AFM} (b) ($d = 0.61\,{\rm nm}$)
should be a valuable precursor to
the ``snap-in'' scenario that may also run the trouble of damaging 
the instrument.
(The rumor indeed goes that experienced experimenters are capable 
to ``listen and gauge'' the good or bad shape of their instruments.)

\section{Quantum and other noises}
\label{s:noise}

\subsection{Basic noise spectra}
\label{s:noise-basics}


A ``noisy'' signal is an erratic sequence $u(t)$ of data whose properties
can only be characterized in a statistical sense. For example, the (sliding)
average signal may be defined as
\begin{equation}
\overline{u}(t) = \frac{1}{2\tau} \int\limits_{t - \tau}^{t + \tau}\!{\rm d}t' 
\, u(t')
\label{eq:}
\end{equation}
where $2\tau$ is the averaging time. 
In the following, we assume that this
yields a slowly varying function that can be subtracted from the data,
and 
therefore take $\overline{u}(t) = 0$.
The noise power spectral density is a measure of the fluctuations 
and their temporal correlations, 
as determined by the average Fourier transform (Wiener-Khintchine 
theorem \cite{MandelWolf})
\begin{equation}
S(f; t) = \frac{1}{2\tau} \bigg| 
\int\limits_{t - \tau}^{t + \tau}\!{\rm d}t' 
\, u(t') \, {\rm e}^{ 2\pi {\rm i} f t'} 
\bigg|^2
\label{eq:sliding-FT}
\end{equation}
If $2\tau$ is much longer than the correlation time of the signal fluctuations,
i.e., the inverse frequency width of $S(f; t)$, then the spectral density
converges to a quantity independent of $\tau$. 
We then get for the signal variance the Fourier representation
\begin{equation}
\overline{u^2}(t) = \int\!{\rm d}f \, S(f; t) 
\label{eq:}
\end{equation}
The quantity $[S(f; t) \, {\rm d}f]^{1/2}$ thus gives the rms signal amplitude
in a bandwidth ${\rm d}f$ around $f$. This suggests the following synthesis
of a noise signal
\begin{equation}
u(t) = \sum_f [S(f; t) \, {\rm d}f]^{1/2} \, {\rm e}^{ {\rm i}\varphi(f) - 
2\pi {\rm i} f t} 
\label{eq:synthetic-noise}
\end{equation}
where $f$ is sampled with spacing ${\rm d}f$, and 
$\varphi(f)$ is a randomly chosen phase
\ch{(see also Ref.\:\cite{Blencowe25})}.
Alternatively, we may draw a random amplitude $A(f)$ 
from a complex normal distribution with variance $S(f; t) \, {\rm d}f$.
To synthesize a stereo signal, real and imaginary parts
of $u(t)$ may be used for the left and right channels.

\subsection{Examples}
\label{s:noise-examples}

\begin{figure}[bthp]
   \centering
   \includegraphics[width=0.45\textwidth]{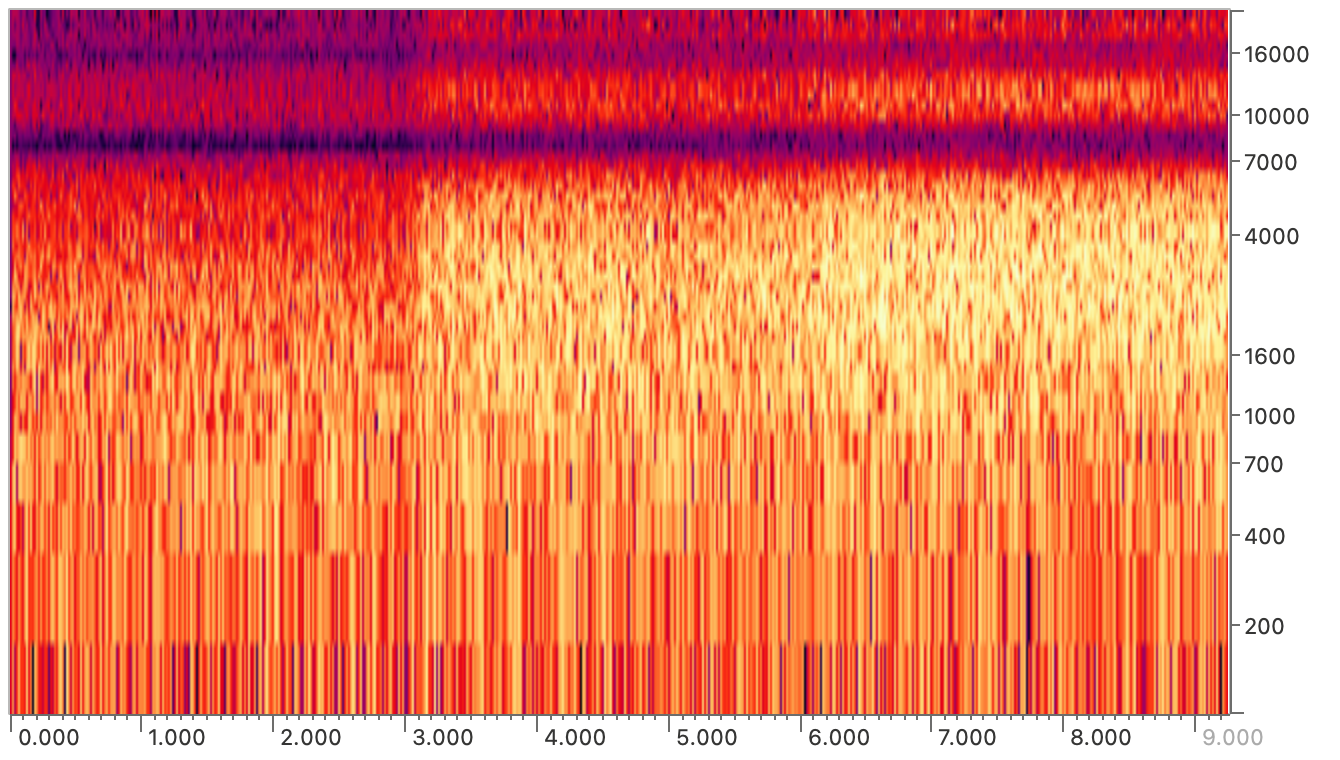}
   \caption[]{Synthetic noise with a thermal spectrum (\ch{first 3.25s}), 
   including quantum noise and a white spectrum (\ch{last 3.25s}). 
   The dip around 8\,kHz in the spectrogram is probably due to the digital 
   sampling.}
   \label{fig:three-noises}
\end{figure}

The audio file 
\linkAudio{noise\_thermal\_quantum\_white.mp3}
(see also Fig.\:\ref{fig:three-noises})
contains three tracks of noise, computed from
different spectra according to Eq.\,(\ref{eq:synthetic-noise}).
The first is a classical thermal spectrum 
with a Drude cutoff at high frequencies, 
the second is quantum noise with a symmetrized spectrum,
the third one is purely white:
\begin{eqnarray}
S_T(f) &=& \frac{k T/\tau}{1 + (f \tau)^2 }\,,
\nonumber\\
S_q(f) &=& \frac{h f}{2\tau} \frac{ \coth hf/(2kT) }{ 1 + (f \tau)^2 }\,,
\nonumber\\
S_w(f) &=& \text{const.}
\label{eq:}
\end{eqnarray}
We take the relatively arbitrary parameters $kT = 3 h f_c$,
$1/\tau = 10\,f_c$ and map the frequency range 
$0 \ldots f_{\rm max} = 50\,f_c$ linearly onto the audible interval 
$164 \ldots 8192\,{\rm Hz}$. The amplitudes of the spectra \ch{are}
normalized to a common value at $\approx 325\,{\rm Hz}$.
The main acoustic impression is that lower frequencies dominate the
thermal spectrum. 
Quantum and white noise sound more harsh or ``metallic'', 
and are more difficult to tell apart.%
\footnote{%
White noise is actually an artefact since it has infinite
energy when integrated over the spectrum. 
The eigenfrequency spectrum of any physical system should have 
an upper cutoff, probably even field theories when the Planck scale
is considered.
}

\begin{figure}[tbph]
\centering
\includegraphics[width=07cm]{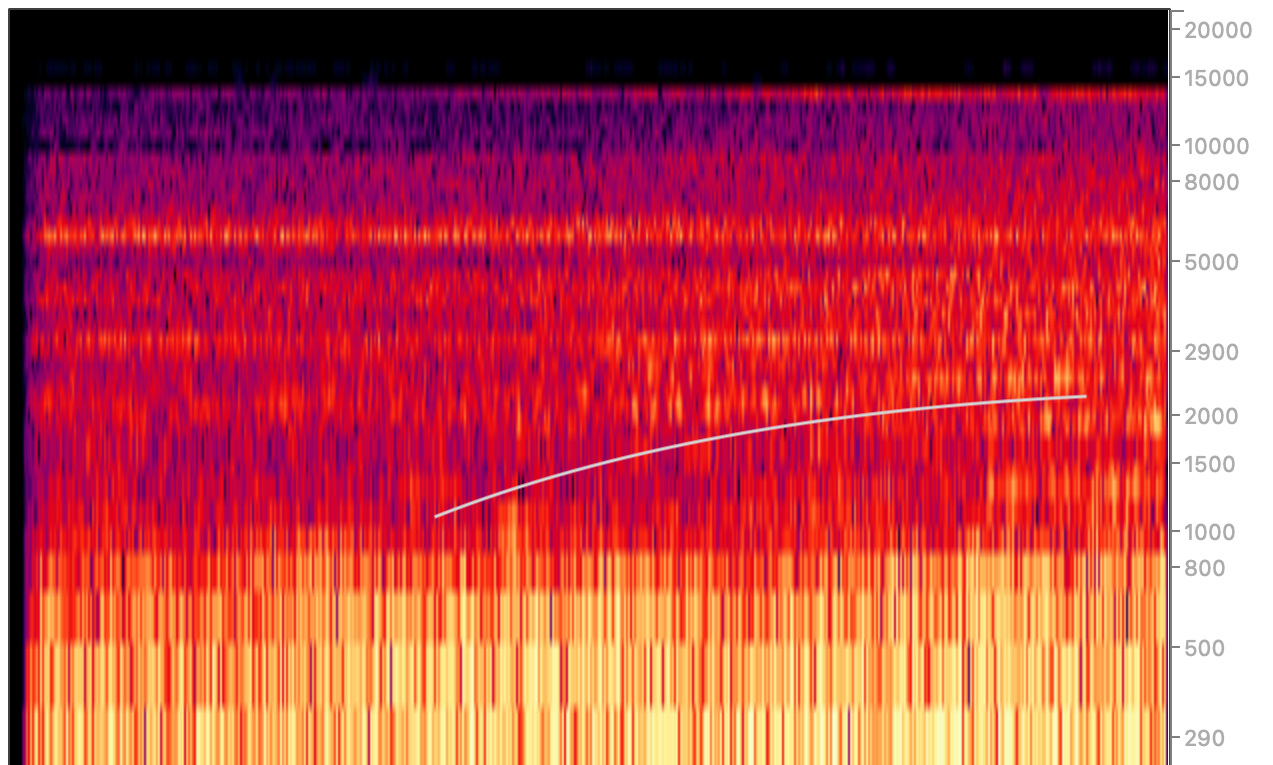}
\caption[]{Spectrogram for a simple recording of a railway engine 
revving up.
Logarithmic scale for frequency axis. 
A close-to-diatonic scale climbing up can be faintly seen (thin gray
line to guide the eye), but is clearly discernible in the audio file 
\linkAudio{locomotive\_rev\_up.mp3}. 
The duration shown is $4\,{\rm sec}$.}
\label{fig:locomotive}
\end{figure}

A mixed time and frequency representation is called a spectrogram and displays
as a function of time a sliding Fourier transform similar to
Eq.\,(\ref{eq:sliding-FT}) (thus providing information like
a musical score). 
Different temporal
filters for the time window are available (rectangle, Hamming, Blackman-Nuttall 
etc.). Examples are given in Figs.\:\ref{fig:three-noises}, \ref{fig:locomotive}.
The first figure displays the result of audio software
analyzing the synthetic noise samples, and one sees the stronger
contribution of higher frequencies, as the noise gets ``quantum'' or ``white''.
In the second example, 
we hear a locomotive gearing up, following nearly a diatonic scale. 
This ``musical'' signal is superimposed on broad-band noise.

\subsection{Local energy spectrum of a Bose gas}
\label{s:Bose-gas-spectrogram}

\begin{figure}[htbp]
   \centering
   \includegraphics[height=0.13\textwidth]{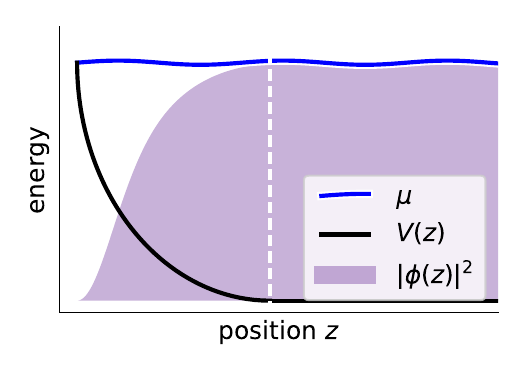}
   \hspace*{03mm}
   \includegraphics[height=0.13\textwidth]{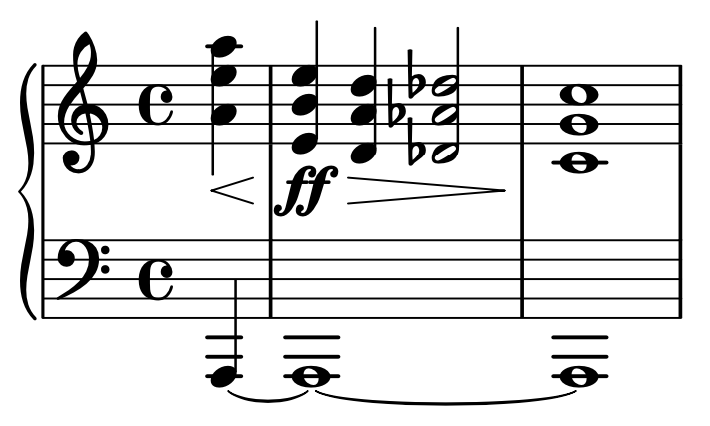}
   \includegraphics[width=0.47\textwidth]{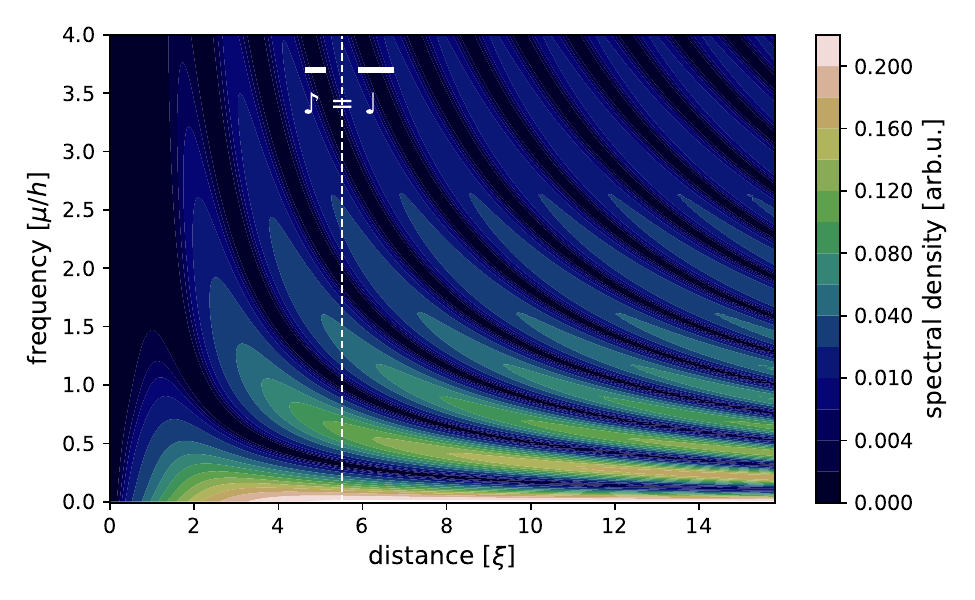} 
   \caption[]{Local spectrum of energy fluctuations in the ground
   state of a Bose condensate. The system is confined to the half-space
   $z \ge 0$ by a ``quarter-pipe'' potential (top left panel).
   The top right panel is a crude musical score (\linkAudio{sonify\_deep\_a.mp3}): 
   it tries to capture
   the condensate as a base tone,
   the lowering of the frequency spectrum and its slowing down, as
   one moves away from the edge position $z = 0$. 
   Distances scaled to the healing length $\xi = \hbar/(m \mu)^{1/2}$
   with the particle mass $m$ and the chemical potential $\mu$,
   typical values in Rb-87 
   are \ch{$\mu/h = 2100\,{\rm Hz}$} and $\xi = 230\,{\rm nm}$.
   --
   For the audio file
   \linkAudio{quantum\_beach.mp3}, the spectrum plotted in the lower panel 
   is mapped into
   the audio range, and computed for a few positions separated by the
	horizontal bar lines. At the dashed line (where
   the potential $V(z)$ joins a flat bottom $V = 0$), the spacing
   is doubled (similar to a dub-step indication).}
   \label{fig:Bose-score}
\end{figure}

We have discussed in a previous paper \cite{Henkel25b_arxiv}
inhomogeneous Bose gases and their local quantum energy density.
In a typical setting, a system of a few
thousand atoms is confined to a thin quasi-one-dimensional
geometry, closed on one end by a potential $V(z)$.
The spectrum of the energy density has been computed in the
degenerate limit where most of the particles reside in the
condensate mode $\phi(z)$. 
In the jargon of quantum field theory for the grand-canonical
ensemble, this would be called the ``vacuum state''. 
Due to interactions, however,
a nonzero fraction of particles appears in excited modes
denoted $v(z; f)$, even at zero temperature (``quantum fluctuations'').
The frequencies $f$ form a continuous spectrum,
unless the system is closed by confinement at the other end.
It has been shown that the quantum fluctuation energy 
in bandwidth ${\rm d}f$ is given by
\cite{AlKhawaja02b, Mora03}
\begin{equation}
S(z, f) \, {\rm d}z\,{\rm d}f = f \, |v(z; f)|^2 \, {\rm d}z\,{\rm d}f
\label{eq:}
\end{equation}
provided the $v(z; f)$ are suitably normalized (see \cite{Henkel25b_arxiv}).
In Fig.\:\ref{fig:Bose-score}, the result of a numerical calculation
is shown as a spectrogram in the $zf$-plane and, in a rough approximation, 
as a musical score. The confining potential is sketched in the upper left panel.
The oscillations represent nodal lines because the modes $v(z; f)$
form standing waves at some effective potential barrier.

The sonification of these data poses the challenge of
shifting the spectrum to some suitable reference frequency. 
Indeed,
the frequency $f$ is actually measured relative to the so-called chemical
potential $\mu$ which gives the ``vibration'' of the condensate mode.
(It is visible as the spectral weight near $f = 0$ in 
Fig.\:\ref{fig:Bose-score}, lower panel.)
An oscillation at $\mu/h$ is,
however, not absolutely measurable because 
there is a phase (or U(1)) symmetry in quantum mechanics inherited 
from the arbitrary choice of a reference energy in classical mechanics.
The situation is similar to transmitting an audio signal by frequency
or amplitude modulation over the radio: the carrier frequency is irrelevant
for the acoustic spectrum.
We choose an algorithm similar to the noise spectra of Fig.\:\ref{fig:three-noises}
and map the range $\mu + h f = \mu \ldots 67\,\mu/h$ to the acoustic range
$82 \ldots 4096\,{\rm Hz}$, using Eq.\,(\ref{eq:synthetic-noise}) and
the fast Fourier transform.
The condensate which would appear with a sharply
defined frequency, is not included.
%

The na\"{\i}ve musical score shown in the top right panel of 
Fig.\:\ref{fig:Bose-score}
is based on the observation
that the spectrum is dominated, at a fixed position, by a few bands with
simple frequency ratios, hence the chords. This expectation is not
borne out by the actual synthesized sound: it is far from musical and
rather a rough reminder of aircraft noise. The frequency bands are probably
too wide to allow for perceiving a collection of individual tones.

\begin{figure}[htbp]
   \centering
   \includegraphics[width=0.37\textwidth]{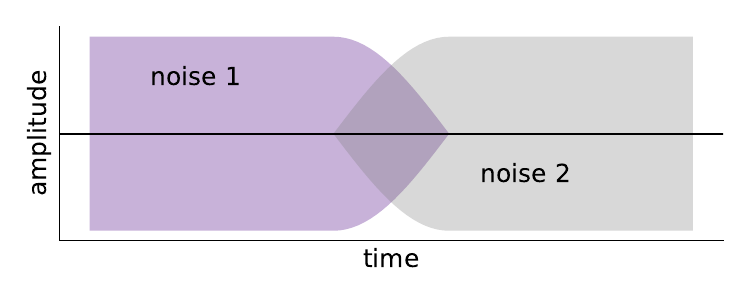} 
   \caption[]{Mixing two noise signals while keeping constant the 
   overall loudness impression (which is quadratic in the signal amplitude).
   In the overlap region of the two noise samples, the signals
   are multiplied with cosine and sine functions
   whose squares sum to unity, and then added.}
   \label{fig:ueberblende}
\end{figure}

To produce a smooth ``promenade'' across the spatial coordinate, we have
patched together noise signals of 1/3\,sec each, using a mixing algorithm
that weights the signal envelopes such their squared sum stays constant, 
as sketched in Fig.\:\ref{fig:ueberblende}. This blending works well for
noise signals and avoids
knack artefacts from phase slips and other discontinuities.

%
%
%
%


\section{Conclusion}
\label{s:conclusion}

A quantum physicist with some affinities to noise and music has tried 
himself for this paper in \sout{amateur}sonification. In a few examples, 
I have illustrated or better: brought to the ear of the reader,
how visual and audio representations are different.
The two main human perceptions indeed complement themselves quite
efficiently when they are confronted with a complex signal, 
\ch{as also observed by Iannis Xenakis:

``We are capable of speaking two languages at the same time. One is addressed to the eyes, the other to the ears. The content of the communication is different but sometimes there's a link between the two.''
\cite{wiki_Xenakis}
}

The ear is a superb Fourier transformer and filters even weak signals
out of ambient noise. A spectrum that looks clustered becomes
more evenly spaced in the logarithmic scale familiar from musical
notation (Fig.\:\ref{fig:hydrogen-piano}).
The eye captures information at a glance, but both senses 
require some training and background culture \sout{to be able}to identify
features. The mathematical music pieces sketched here
may be fun to train your acoustic memory. An additional
\textbf{exercise} inspired by number theory would be to use the (approximately
logarithmic) distribution of prime numbers as a source of ``inspired
randomness'' for synthesizing a sequence of notes. 
Finally, the examples from atomic physics have brought forward the
unusual separation of time scales related to high quality
factors (Sec.\:\ref{s:quantum-chord}) -- or conversely the fairly
broad spectral structures in the local density of quantum energy
fluctuations (Sec.\:\ref{s:Bose-gas-spectrogram}). 

As a final \textbf{exercise}, I would suggest to sonify the jump dynamics
of an individual quantum state 
\cite{Sauter1986, Stenholm1997}
and encode it in a short list of 
pitches (its spectrum)
whose intensities are proportional to the current
probability amplitudes. 
\ch{Stereo signals can provide a way to convey quantum coherences 
(relative phases) \cite{Christie24}.}
This way of ``listening to the quantum''
will certainly not resolve the discussion around the meaning
of a quantum measurement \cite{Mermin2014}, but perhaps provide
a complementary challenge to the participants' senses.

%

\vskip 2ex

\acknowledgement{%
\textit{Acknowledgements.}
The graphics in this paper have been produced with the Matplotlib
and Ocenaudio software, the musical scores with Lilypond,
digits of $\pi$ and $e$ were provided by Wolfram Alpha.
The research of C.H.\ is supported by the 
\emph{Deutsche Forschungsgemeinschaft} (DFG) with SFB1636, 
project ID 510943930, Projects No.\ A01 and A04. 
I thank Regina Hoffmann-Vogel for shared lecturing on force
microscopy.
}


\appendix

\section{Appendix}
\label{a:other-examples}

\begin{figure}[tbph]
\centering
\includegraphics[width=08cm]{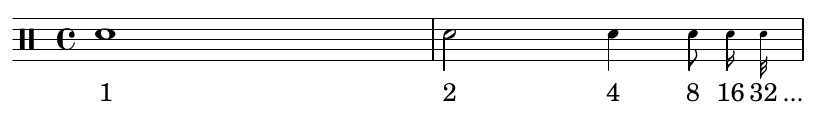}
\caption[]{Geometric series of infinitely many notes, 
summing up to two bars. Audio file: \linkAudio{geometric\_rhythm.mp3}.}
\label{fig:geometric}
\end{figure}

A few other illustrations (and audio samples) are given
here. A geometric series whose infinitely many terms sum up
to a finite number can be represented as an infinite accelerating
rhythm (Fig.\:\ref{fig:geometric})
or simply as a ``geometric typography'', e.g.:
\begin{equation}
2^{1/12} = 
\def\defaultstartht{24pt}
\diminish[0.9438744]{\well}
\label{eq:well-tempered}
\end{equation}

%


%
\begin{figure*}[htbp]
   \centering
   \includegraphics[width=13cm,clip]{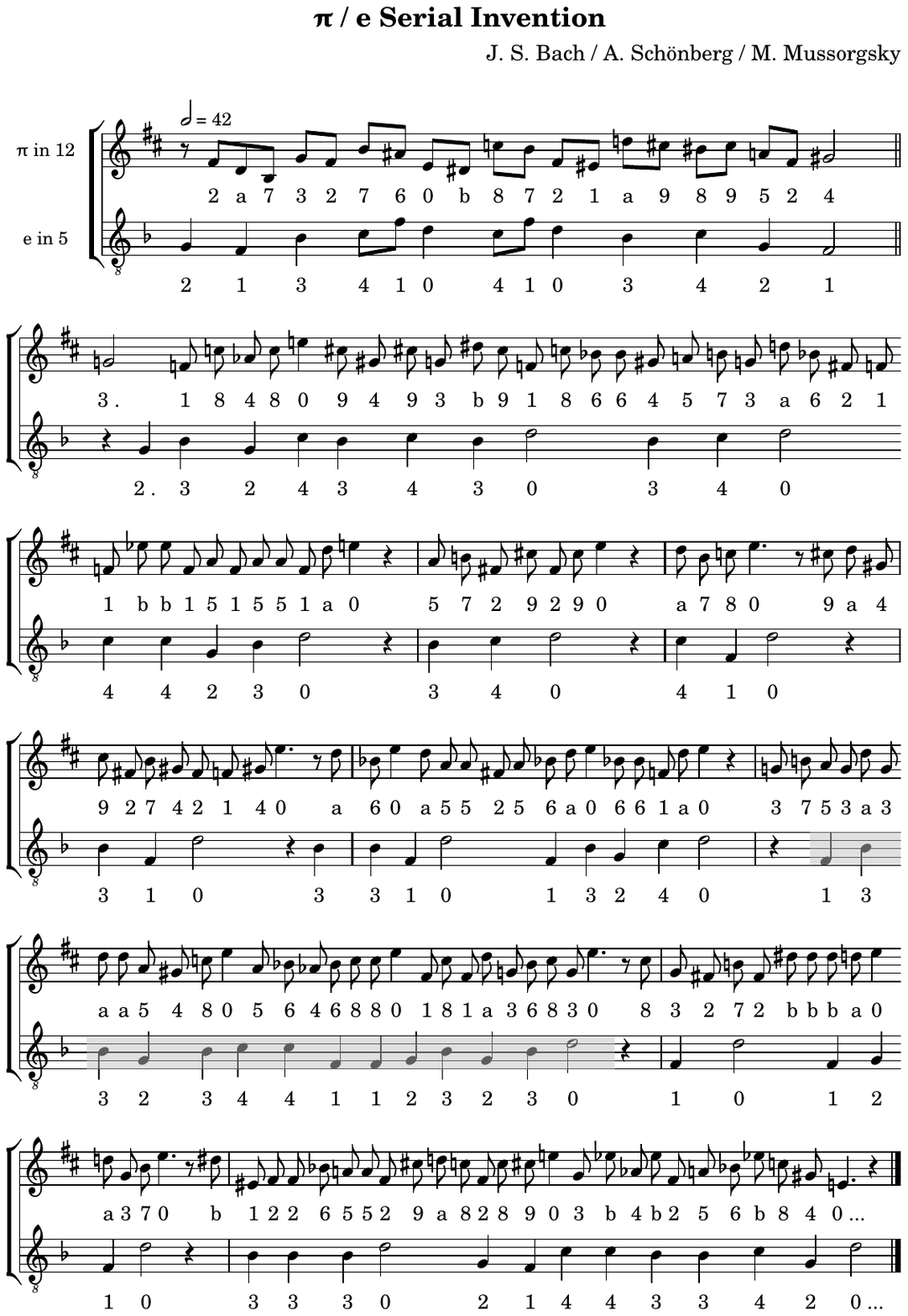}
   \caption[]{Full $\pi/{\rm e}$ sonification sketched in
   Fig.\:\ref{fig:pi-over-e-thumb}:
   starting from the double bar, duet between an 
   upper chromatic voice based on 
   141 digits of $\pi$ to base 12,
   lower pentatonic voice coded by
   67 digits of ${\rm e}$ to base 5.
   Introduction: upper voice = theme of Fuga no.~24 in b minor 
   from the ``Well-tempered Clavier~I'' 
   by J.\,S.\ Bach, a dodecaphonic series with a few repeated notes;
   lower voice = Promenade motif from
   ``Pictures of an Exhibition'' by M.\ Mussorgsky. 
   The juxtaposition of the two keys (b minor and B$\flat$ major) 
   is done disregarding
   any rules of musical composition.
   The lower-voice passage marked in grey illustrates a sequence 
   that may be perceived as musically ``meaningful''.
   A few enharmonic substitutions are applied 
   to ease the music reading. 
   The digit zero is emphasized with the double duration, and the
   music briefly halts when the voices reach 0 simultaneously.  
   Audio file
   \linkAudio{pi\_over\_e\_classical.mp3}, 1\,min 14\,sec long (slow tempo).}
   \label{fig:pi-over-e}
\end{figure*}

\begin{figure*}[htbp]
   \centering
   \includegraphics[width=13cm]{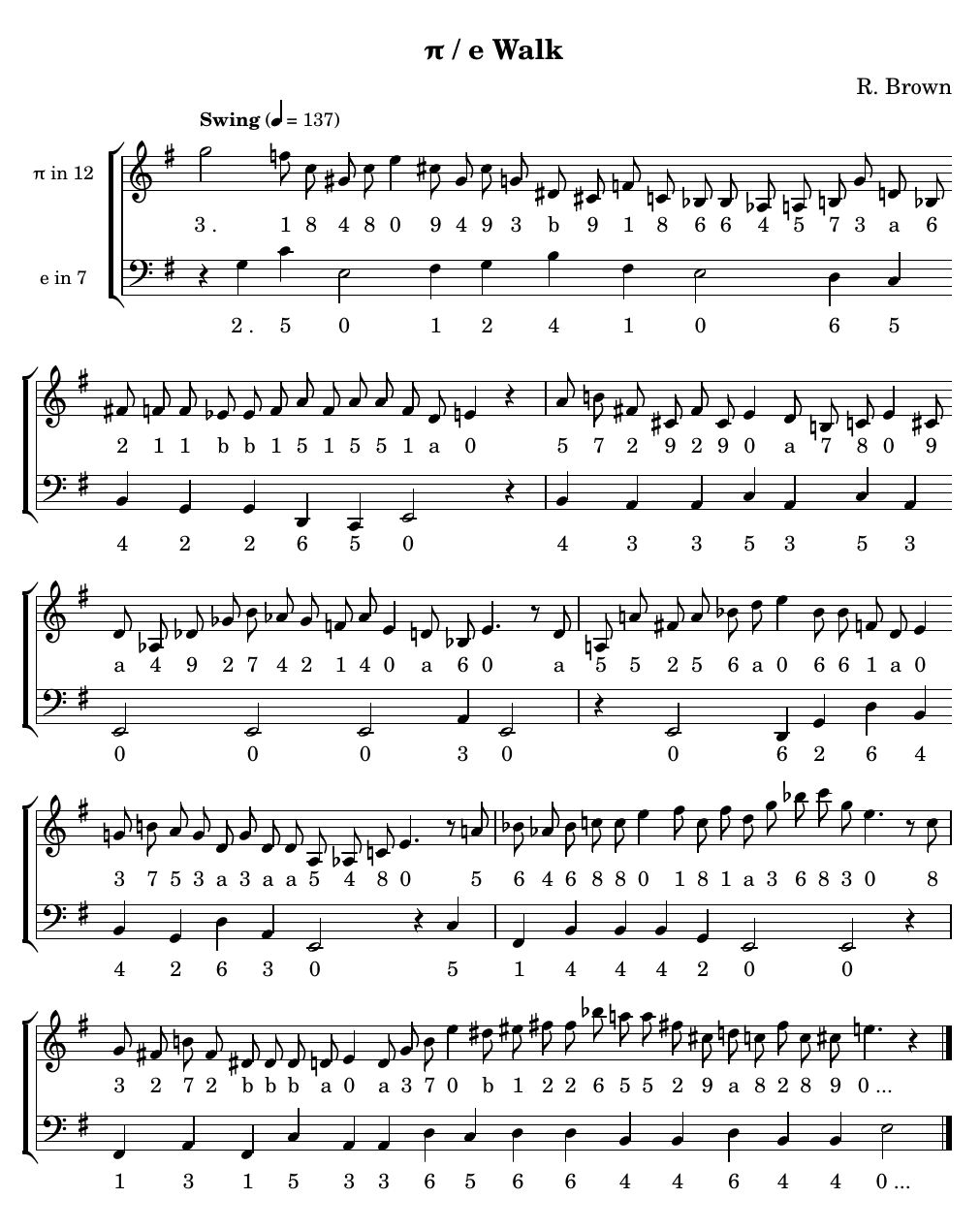} 
   \caption[]{Brownian motion rendering of 
   $\pi/{\rm e}$: same 12-digit encoding as in 
   Fig.\:\ref{fig:pi-over-e} for the $\pi$ voice, 
   but a base-7 encoding to a diatonic scale (e minor) for the lower e voice.
   The Lilypond command \texttt{\small $\backslash$relative}
   has been used that takes the smallest interval (up or down)
   between two notes. A few adjustments have been made to remain
   within a reasonable ambitus for the two voices;
   digit 0 notes with double duration, brief halt when both voices reach 0.
   Audio file \linkAudio{pi\_over\_e\_Walk\_127.mp3}, 37\,sec long in the
   given tempo, with some swing feel rendering.
   }
   \label{fig:pi-over-e-walk}
\end{figure*}

\begin{figure*}[htbp]
   \centering
   \includegraphics[width=13cm]{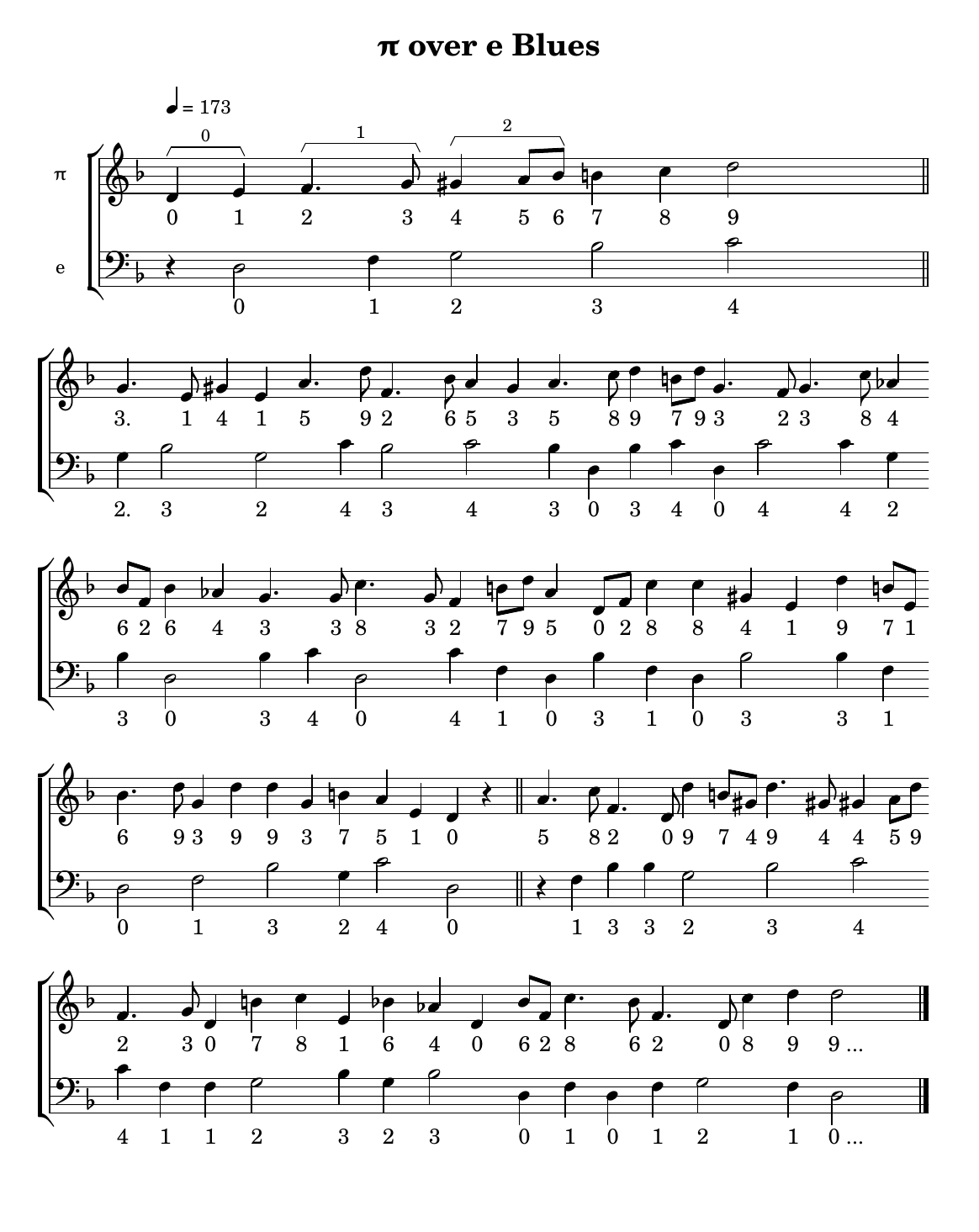} 
   \caption[]{$\pi/{\rm e}$ with decimal digits in the $\pi$ voice,
   base 5 in voice e.
   A blues-like diatonic scale based on D (ten notes in total) 
   has been combined with the
   pentatonic scale of Fig.\:\ref{fig:pi-over-e}, lower voice. 
   The digits of the ratio $\pi/e 
   = 1.1557273498 \ldots$
   in base 3 and base 2 are specifying the rhythmic patterns of the
   two voices. 
   They are presented in the first line
   together with the pitch-to-digit code 
   (upper brackets for three basic patterns). 
   The rhythm of the lower voice is binary with
   $0$\,: half note, $1$\,: quarter note. 
   One brief halt when both voices reach digit 0.
   Audio file \linkAudio{pi\_over\_e\_Blues.mp3}, 33\,sec long in the
   given tempo.
   }
   \label{fig:pi-over-e-Blues}
\end{figure*}
\end{document}